\documentclass[12pt]{article}
\usepackage{amsmath}
\usepackage{amsbsy}
\usepackage{amssymb}
\usepackage[all]{xy}
\usepackage{xcolor}
\usepackage{float}
\usepackage{fancyvrb}
\PassOptionsToPackage{hyphens}{url}\usepackage{hyperref}
\usepackage{comment}
\usepackage{authblk}

\input{x.cfg}

\title{A Neat Linked Queue with the Rear Blank Node}
\author{Xie Xie \\ \texttt{xiexiexx@gmail.com}}

\affil{\small{}School of Cyberspace Security, \\Xi'an University of Posts and Telecommunications, Xi'an, China}

\begin{document}
\maketitle

\begin{abstract}
We introduce a very simple queue implementation with the singly linked list. With the help of the rear blank node instead of the usual header node, we avoid additional check steps for the dequeue operation in the traditional implementations existing for many decades. The essence of our representation is the half-opened pointer interval with the same direction of the queue operations, which can guarantee the uniform treatment even in the empty queue case. The simplification of queue implementations cuts off unnecessary steps, and it minimizes the number of steps in the dequeue operation with the time limitation of enqueue operation, which could contribute to the performance of the real-time systems. We extend the linked queue to the circularly linked queue, which can also be used to implement stack and take advantage of the maximal information of the single direction in the circularly linked list, and it actually constructs the output-restricted deque. We also present a variant: lazy circularly linked queue, which is more efficient in some special cases, especially for the dequeue operations.
\end{abstract}


\section{Introduction}

Queue is one of the most important data structure in computer science, which only has two basic operations: \texttt{enqueue} and \texttt{dequeue}. There are various implementations for queue, such as the class template \texttt{queue} based on dynamics arrays in \CC{} Standard Template Library (STL). 

For real-time systems, we need the reliable queues with $O(1)$ time operations in the worst case. The doubly linked list can guarantee such time bound, but it stores two link fields. We may use simpler data structures, such as the singly linked list, but it has an unavoidable problem of additional check steps in the \texttt{dequeue} operation, as mentioned by Knuth in the pages 260-261 of his classical book \cite{Knuth-v1}:
\begin{quote}
We will make use of pointers \texttt{F} and \texttt{R}, to the front and rear (of the queue) ... 

Notice that \texttt{R} must be changed when the queue becomes empty; this is precisely the type of ``boundary condition'' we should always be watching for.
\end{quote}

This problem has been existing for many decades, and nowadays it is more obvious to the high-throughput queue.
For example, if a queue deals with one billion dequeue operations in 30 seconds,
at least two billion instructions (\texttt{cmp} and \texttt{jne}) are totally unnecessary.

We will present a simple way to solve the problem in the singly linked list for implementing the queue, and these methods are now included into \emph{The Art of Computer Programming} as two exercises.\footnote{\url{https://www-cs-faculty.stanford.edu/~knuth/err1.ps.gz}, exercise 30 in section 2.2.3 (17 Aug 2021) and exercise 4 in section 2.2.4 (16 Aug 2021).}

\section{Data Structures}

We use the \CC{} language to implement the singly linked list in the form of generic programming, whose nodes type \texttt{lnode} is defined as:
\begin{Verbatim}[frame=leftline]
template <typename T>
struct lnode {
  T data;
  lnode<T>* next;
};
\end{Verbatim}
in which \texttt{T} is the type of data stored in the list. The pointers in this paper are all with the type \texttt{lnode<T>*}. 
Based on the strategy of STL, we don't use the \texttt{dequeue} operation when a queue is empty, which means the \texttt{dequeue} operation will call an emptiness check function in advance.

A queue can be represented by a sequence. For example, the queue $Q = (0, 1, 2)$ stored 3 elements, whose front element is 0 and rear element is 2.

\section{Direct Method}

\begin{figure}[ht]
\begin{equation*}
\xymatrix@C=0pt@R=12pt
{
                                                                                                                              \mPL{H}\ar[d]                                                   &                                       &       & \mPL{F}\ar@{.>}[d]                                                   &                                          &    & & &  & \mPL{R}\ar[d] \\
\xANode{\color{white}\mPL{00000}} & \xLNode\save *{\bullet}\ar"2,4"\restore & \quad & \xANode{{\color{white}\mPL{00}}\mPL{0}{\color{white}\mPL{00}}}\save+<0pc, -1.5pc>*{\mPL{front}}\restore & \xLNode\save *{\bullet}\ar"2,7"\restore & \quad & \xANode{{\color{white}\mPL{00}}\mPL{1}{\color{white}\mPL{00}}} & \xLNode\save *{\bullet}\ar"2,10"\restore & \quad & \xANode{{\color{white}\mPL{00}}\mPL{2}{\color{white}\mPL{00}}}\save+<0pc, -1.5pc>*{\mPL{rear}}\restore & \xLNode\save *{\circ}\restore\\
}
\end{equation*}
    \caption{Singly linked queue}
    \label{singly_linked_queue}
\end{figure}
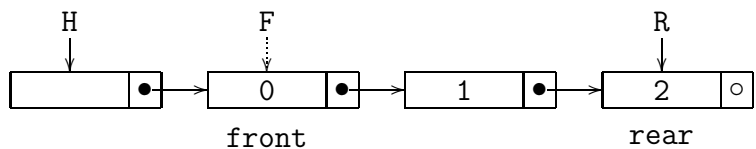

The header (or dummy node) with the pointer \texttt{H} which sits before the first element in the singly linked queue is usually set to avoid the empty case. In fact, the pointer \texttt{H->next} plays the role of the pointer {F}. Figure \ref{singly_linked_queue} gives an queue example $Q = (0, 1, 2)$, and the \texttt{NULL} pointer is denoted by the symbol ``$\circ$''.

When we dequeue an element from the queue, we need to move \texttt{H->next} to the next node and free the original node. But \texttt{R} must be reset to \texttt{H} since the node pointed by \texttt{R} may be freed. It is not a convenient solution, and the ``boundary condition'' is still needs to be monitored. 

The code of \texttt{dequeue} is:
\begin{Verbatim}[frame=leftline]
template <typename T>
void dequeue()
{
  lnode<T>* F = H->next;
  H->next = F->next;
  if (F == R)     // additional checking.
    R = H;        // additional assignment.
  delete F;
}
\end{Verbatim}
and it has the additional check steps.

The code of \texttt{enqueue} is:
\begin{Verbatim}[frame=leftline]
template <typename T>
void enqueue(const T& item)
{
  lnode<T>* p = new lnode<T>;
  p->data = item;
  p->next = NULL;
  R->next = p;
  R = p;
}
\end{Verbatim}
in which \texttt{item} is the element to be enqueued.

Why we need to check the value of \texttt{R} when each time we dequeue an element of the queue? The essence of the problem is the queue description.

\section{Interval}

The problem of the traditional implementations lies in the representation of queue. It is worth noting that we often use a closed pointer interval to represent the linked queue, i. e. $[\mbox{\texttt{F}}, \mbox{\texttt{R}}]$. 
The key of the operations in the queue is the movement, which is similar to the successor function in the Peano axiom. Since the singly linked list only has one direction, we can define the successor function $\mathcal{S}$ as:
\begin{equation*}
\mathcal{S}(\mbox{\texttt{p}}) = \mbox{\texttt{p->next}},
\end{equation*}
in which \texttt{p} is a pointer to a node in the queue.
Then the pointer interval for the queue $Q$ in the Figure \ref{singly_linked_queue} is actually
\begin{equation*}
[\mbox{\texttt{F}}, \mbox{\texttt{R}}] = \big(\mbox{\texttt{F}}, \mathcal{S}(\mbox{\texttt{F}}), \mathcal{S}(\mathcal{S}(\mbox{\texttt{F}}))\big).
\end{equation*}
In fact, both of the \texttt{enqueue} and \texttt{dequeue} use the successor function to deal with the singly linked list.

We only know one direction in the singly linked list with the \texttt{next} pointer field, and the singly linked list can only delete the elements after the current position in $O(1)$ time without the information of other nodes, thus the \texttt{dequeue} operation must deal with the leftmost of the interval, and the interval direction must be from \texttt{F} to \texttt{R}. 

Ideally, we only change \texttt{F} in the \texttt{dequeue} operation and \texttt{R} in the \texttt{enqueue} operation. 
When $Q$ only has one element, the interval will be 
\begin{equation}
[\mbox{\texttt{F}}, \mbox{\texttt{R}}] = [\mbox{\texttt{e}}, \mbox{\texttt{e}}],
\end{equation}
in which \texttt{e} is the pointer value of \texttt{F} and $\mathcal{S}(\mbox{\texttt{e}})$ is \texttt{NULL}. 
If we free the last remaining element from the queue, 
the interval will be 
\begin{equation}
[\mbox{\texttt{F}}, \mbox{\texttt{R}}] = [\mathcal{S}(\mbox{\texttt{e}}), \mbox{\texttt{e}}] = [\mbox{\texttt{NULL}}, \mbox{\texttt{e}}].
\end{equation}
and the pointer \texttt{R} will be invalid (since the node pointed by \texttt{e} is freed). Thus, we need to maintain both \texttt{F} and \texttt{R} in the \texttt{dequeue} operation.

The method with header uses a half-opened pointer interval $(\mbox{\texttt{H}}, \mbox{\texttt{R}}]$, and the header node position only handles the empty case in the \texttt{enqueue} operation. \texttt{H} is actually the position previous to \texttt{F}, but the \texttt{dequeue} is the forward movement, thus \texttt{H} has no help with the \texttt{dequeue} operation of a queue, and we need the next position to \texttt{R}.

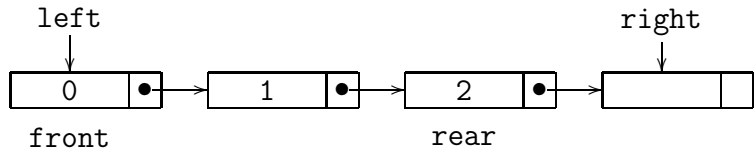
\begin{figure}[ht]
\begin{equation*}
\xymatrix@C=0pt@R=12pt
{
                                                                                                                              \mPL{left}\ar[d]                                                   &                                       &       &                                                    &                                          &    & & &  & \mPL{right}\ar[d] \\
\xANode{{\color{white}\mPL{00}}\mPL{0}{\color{white}\mPL{00}}}\save+<0pc, -1.5pc>*{\mPL{front}}\restore & \xLNode\save *{\bullet}\ar"2,4"\restore  & \quad & \xANode{{\color{white}\mPL{00}}\mPL{1}{\color{white}\mPL{00}}} & \xLNode\save *{\bullet}\ar"2,7"\restore & \quad & \xANode{{\color{white}\mPL{00}}\mPL{2}{\color{white}\mPL{00}}}\save+<0pc, -1.5pc>*{\mPL{rear}}\restore & \xLNode\save *{\bullet}\ar"2,10"\restore & \quad & \xANode{\color{white}\mPL{00000}} & \xLNode\save *{}\restore\\
}
\end{equation*}
    \caption{Rear blank node (pointed by \texttt{right})}
    \label{rear_blank_node}
\end{figure}

The idea of queue in the circular array can also inspire us. The front of the queue is recorded, and the next position of the rear of the queue is indicated. 
We can describe a queue with a half-opened interval 
$[\mbox{\texttt{left}}, \mbox{\texttt{right}})$, in which \texttt{left} denotes the pointer to the front of the queue and \texttt{right} denotes the pointer to a blank node (we can call it the \emph{rear blank node}) after the rear of the queue. It is noticed that the rear blank node do not store any real elements in the queue and such type of half-opened interval can properly describle the direction the queue operations. Furthermore, the pointer \texttt{right} itself plays the role of the \texttt{NULL} pointer, thus we 
no longer need \texttt{NULL} in such a queue. Figure \ref{rear_blank_node} shows an example of the queue $Q = (0, 1, 2)$.

\section{Blank Node Reserved The Space}

The blank node pointed by \texttt{right} which sits after the rear of queue can solve our problem, since the rear blank node can reserve the space for the effects of successor function in the \texttt{dequeue} operation. In fact, the difference between a header and such rear blank node is: \emph{the rear blank node is always brand-new when an element is enqueued}. Now the operations of the queue are very simple, especially in the \texttt{dequeue} operation:

\begin{Verbatim}[frame=leftline]
template <typename T>
void dequeue()
{
  lnode<T>* p = left;
  left = left->next;
  delete p;
}
\end{Verbatim}

\begin{Verbatim}[frame=leftline]
template <typename T>
void enqueue(const T& item)
{
  right->data = item;
  lnode<T>* p = new lnode<T>;
  right->next = p;
  right = p;
}
\end{Verbatim}

We no longer need to check the pointer \texttt{right} in the dequeue operation, since \texttt{right} always exists in the whole lifetime of the queue and reserve the space for the $\mathcal{S}(\mbox{\texttt{left}})$. It is noticed that the newly enqueued element is assigned to the old rear blank node, then we add a brand-new rear blank node.

 Since the \texttt{enqueue} operation involves compound item assignment, it needs to run fast in the real-time system, and this time limitation of the \texttt{enqueue} operation should not be exceeded in the other operations. Our method minimize the number of steps in the \texttt{dequeue} operation with the time limitation of \texttt{enqueue} operation. The statuses (data and pointer) of the involved nodes in the queue need to be updated:
\begin{itemize}
    \item   \texttt{enqueue}: In the worst case, a new node needs to be allocated and a new item needs to be assigned to somewhere in the queue, then the value of \texttt{right->next} (for linking) and \texttt{right} (for indicating the status for the next linking) must be changed. It is noticed that we do not need to assign \texttt{NULL} to \texttt{p->next}. Thus we at least need 4 steps to construct the \texttt{enqueue} operation.
    \item   \texttt{dequeue}: The value of \texttt{left} must be changed for indicating the new front element, but the original value of \texttt{left} needs to be saved to \texttt{p} in advance, and finally the old front node is freed. If we do not free the original node, we can save it to somewhere temporarily (such as a private list only for us) for the future. But such reserved nodes will be used up, so we will check them in the \texttt{enqueue} operation, but it is not allowed for the time limitation. Thus we at least need 3 steps to construct the \texttt{dequeue} operation.
\end{itemize}
It can be concluded that \textit{our implementations have the minimum number of assignments and have not any redundant steps in the case of being with the time limitation}.

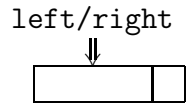
\begin{figure}[ht]
\begin{equation*}
\xymatrix@C=0pt@R=12pt
{
                                                                                                                              \mPL{}\ar@{=>}[d]\save+<0pc, 0.2pc>*{\mPL{left/right}}\restore                                                    \\
\xANode{\color{white}\mPL{00000}} & \xLNode\save *{}\restore\\
}
\end{equation*}
    \caption{An empty linked queue with the rear blank node}
    \label{empty_queue}
\end{figure}

By the way, the empty queue is just the pointer interval
\begin{equation}
[\mbox{\texttt{left}}, \mbox{\texttt{right}}) = [\mbox{\texttt{right}}, \mbox{\texttt{right}}),
\end{equation}
since the interval is not really ``empty'', it is safe. Figure \ref{empty_queue} shows an empty linked queue.

\section{Circularly Linked Queue and Stack}

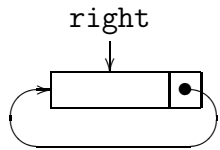
\begin{figure}[ht]
\begin{equation*}
\xymatrix@C=0pt@R=12pt
{
\mPL{right}\ar[d] & \\
\xANode{\color{white}\mPL{00000}} & \xLNode \save *{\bullet}\restore 
\save*{}
    \ar`r[d]+/r1pc/`[d]+/d0pc/
          `[u]+/l5.5pc/`[][l]
\restore \\
 & \\
}
\end{equation*}
    \caption{Empty circularly linked queue}
    \label{empty_circular_queue}
\end{figure}

With the circularly linked list, we can use only one pointer \texttt{right} to handle the queue, which can be represented by the interval $[\mathcal{S}(\mbox{\texttt{right}}), \mbox{\texttt{right}})= [\mbox{\texttt{right->next}}, \mbox{\texttt{right}})$. Figure \ref{circular_queue} gives an example of a circularly linked queue $Q = (0, 1, 2)$, and Figure \ref{empty_circular_queue} shows an empty circularly linked queue.

The operations of queue can be rewritten as:
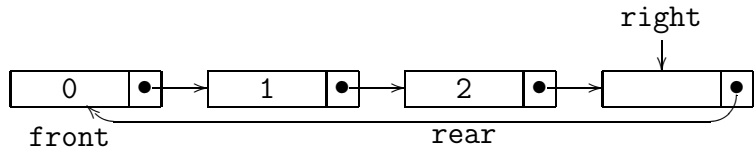
\begin{figure}[ht]
\begin{equation*}
\xymatrix@C=0pt@R=12pt
{
                                                                                                                                                                                 &                                       &       &                                                    &                                          &    & & &  & \mPL{right}\ar[d] \\
\xANode{{\color{white}\mPL{00}}\mPL{0}{\color{white}\mPL{00}}}\save+<0pc, -1.5pc>*{\mPL{front}}\restore & \xLNode\save *{\bullet}\ar"2,4"\restore  & \quad & \xANode{{\color{white}\mPL{00}}\mPL{1}{\color{white}\mPL{00}}} & \xLNode\save *{\bullet}\ar"2,7"\restore & \quad & \xANode{{\color{white}\mPL{00}}\mPL{2}{\color{white}\mPL{00}}}\save+<0pc, -1.5pc>*{\mPL{rear}}\restore & \xLNode\save *{\bullet}\ar"2,10"\restore & \quad & \xANode{\color{white}\mPL{00000}} & \xLNode \save *{\bullet}\ar `d_l"2,1"`_ul"2,1" "2,1"\restore \\
}
\end{equation*}
    \caption{Circularly linked queue}
    \label{circular_queue}
\end{figure}

\begin{Verbatim}[frame=leftline]
template <typename T>
void dequeue()
{
  lnode<T>* left = right->next;
  right->next = left->next;
  delete left;
}
\end{Verbatim}

\begin{Verbatim}[frame=leftline]
template <typename T>
void enqueue(const T& item)
{
  right->data = item;
  lnode<T>* p = new lnode<T>;
  p->next = right->next;
  right->next = p;
  right = p;
}
\end{Verbatim}

The circular structure can also be used to implement the stack. The \texttt{push} and \texttt{pop} operations of the stack are just insert and erase a node after the node pointed by \texttt{right}. Thus the circular structure gives a uniform treatment for the queue and the stack, and it also does not need additional check steps for the empty cases. In fact, the singly linked list constructs the output-restricted deque efficiently (but it can not implement the deque structures), and we take advantage of the maximal information of the single direction with the \texttt{next} pointer field. Furthermore, the circularly linked list now does its best to complete the output-restricted deque: it has a header node when it is a stack, and it has a footer node when it is a queue. By the way, if we are interested only in a stack, the traditional way (such as the implementation in \cite{Knuth-v1}) is easier and faster.

\section{Lazy Circularly Linked Queue}
\label{lazy_section}

If we do not free the dequeued elements during the lifetime of the queue, that will give us a lazy queue, in which the number of the elements is not greater than the capacity of queue. Pointers \texttt{left} and \texttt{right} are both needed in the lazy circularly linked queue. For example, if the queue $Q = (0, 1, 2)$ in Figure \ref{circular_queue} is dequeued without node freeing, it will change to  $Q' = (1, 2)$ in Figure \ref{lazy_circular_queue} with a very simple movement from \texttt{left} to $\mathcal{S}(\mbox{\texttt{left}})$. It is noticed that $Q'$ has 2 elements and the capacity of $Q'$ is still 3.

But the side-effect of the lazy queue is that we must check whether the queue reaches its capacity when a new element is enqueued. Figure \ref{full_lazy_circular_queue} shows such a queue example $Q'' = (1, 2, 3)$, in which \texttt{right->next} is just equal to \texttt{left}.

\begin{figure}[ht]
\begin{equation*}
\xymatrix@C=0pt@R=12pt
{
                                                                                                                             & & & \mPL{left}\ar[d]                                                                                                                                                  &                                          &    & & &  & \mPL{right}\ar[d] \\
\xANode{\color{white}\mPL{00000}} & \xLNode\save *{\bullet}\ar"2,4"\restore  & \quad & \xANode{{\color{white}\mPL{00}}\mPL{1}{\color{white}\mPL{00}}}\save+<0pc, -1.5pc>*{\mPL{front}}\restore & \xLNode\save *{\bullet}\ar"2,7"\restore & \quad & \xANode{{\color{white}\mPL{00}}\mPL{2}{\color{white}\mPL{00}}}\save+<0pc, -1.5pc>*{\mPL{rear}}\restore & \xLNode\save *{\bullet}\ar"2,10"\restore & \quad & \xANode{\color{white}\mPL{00000}} & \xLNode \save *{\bullet}\ar `d_l"2,1"`_ul"2,1" "2,1"\restore \\
}
\end{equation*}
    \caption{Lazy circularly linked queue}
    \label{lazy_circular_queue}
\end{figure}

\begin{figure}[ht]
\begin{equation*}
\xymatrix@C=0pt@R=12pt
{
                                                                                                                             \mPL{right}\ar[d] & & & \mPL{left}\ar[d]                                                                                                                                                  &                                          &    & & &  &  \\
\xANode{\color{white}\mPL{00000}} & \xLNode\save *{\bullet}\ar"2,4"\restore  & \quad & \xANode{{\color{white}\mPL{00}}\mPL{1}{\color{white}\mPL{00}}}\save+<0pc, -1.5pc>*{\mPL{front}}\restore & \xLNode\save *{\bullet}\ar"2,7"\restore & \quad & \xANode{{\color{white}\mPL{00}}\mPL{2}{\color{white}\mPL{00}}} & \xLNode\save *{\bullet}\ar"2,10"\restore & \quad & \xANode{{\color{white}\mPL{00}}\mPL{3}{\color{white}\mPL{00}}}\save+<0pc, -1.5pc>*{\mPL{rear}}\restore & \xLNode \save *{\bullet}\ar `d_l"2,1"`_ul"2,1" "2,1"\restore \\
}
\end{equation*}
    \caption{Lazy circularly linked queue}
    \label{full_lazy_circular_queue}
\end{figure}

The operations in the queue now are:
\begin{Verbatim}[frame=leftline]
template <typename T>
void dequeue()
{
  left = left->next;
}
\end{Verbatim}

\begin{Verbatim}[frame=leftline]
template <typename T>
void enqueue(const T& item)
{
  right->data = item;
  if (right->next == left)
  {
    lnode<T>* p = new lnode<T>;
    p->next = left;
    right->next = p;
    right = p;
  }
  else
    right = right->next;
}
\end{Verbatim}

\begin{figure}[ht]
\begin{equation*}
\xymatrix@C=0pt@R=12pt
{
                                                                                                                             \mPL{}\ar@{=>}[d]\save+<0pc, 0.2pc>*{\mPL{left/right}}\restore & & &                                                                                                                                                   &                                          &    & & &  &  \\
\xANode{\color{white}\mPL{00000}} & \xLNode\save *{\bullet}\ar"2,4"\restore  & \quad & \xANode{\color{white}\mPL{00000}} & \xLNode\save *{\bullet}\ar"2,7"\restore & \quad & \xANode{\color{white}\mPL{00000}} & \xLNode\save *{\bullet}\ar"2,10"\restore & \quad & \xANode{\color{white}\mPL{00000}} & \xLNode \save *{\bullet}\ar `d_l"2,1"`_ul"2,1" "2,1"\restore \\
}
\end{equation*}
    \caption{An empty lazy circularly linked queue}
    \label{empty_lazy_circular_queue}
\end{figure}
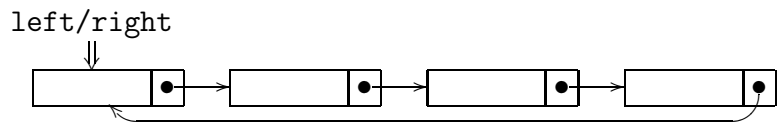

When all the elements in the queue are dequeued, \texttt{left} will be equal to \texttt{right}, and Figure \ref{empty_lazy_circular_queue} shows an empty queue.

If a queue has enough capacity at certain time point and the number of current elements in the queue can be guaranteed not greater than the capacity, the operations of the queue after then will become more efficient, especially the \texttt{dequeue} operations.

\section{Codes}

The full codes can be found at:

\noindent\href{https://github.com/xiexiexx/Puzzles/tree/main/linked-queue-stack}{\url{https://github.com/xiexiexx/Puzzles/tree/main/linked-queue-stack}}.

\section*{Acknowledgements}

The author would like to thank Prof. Knuth for his very precious suggestions and encouragement to this manuscript, and he is so kindly to include the linked queue implementations with the blank node to his masterpiece, \emph{The Art of Computer Programming}. 

\bibliographystyle{alpha}
\newcommand{\etalchar}[1]{$^{#1}$}

\end{document}